\documentclass[graphics]{mn2e}
\newif\ifAMStwofonts \input psfig.sty

\usepackage{rotating}
\include{new_commands}

\def\ao{A0538--66}
\def\til{$\sim$}

\def\deg{$^{\circ}$}

\def\leq{\hbox{${_<\atop{\sim}}$}}
\def\geq{\hbox{${_>\atop{\sim}}$}}

\def\ergsec{\thinspace\hbox{$\hbox{erg}\thinspace\hbox{s}^{-1}$}}

\def\arcsecdot{\nobreak\ifmmode{''\hskip-0.45em.\hskip0.08em}%
                         \else{$''\hskip-0.45em.\hskip0.08em$}\fi}
\def\arcsec{\nobreak\ifmmode{''\hskip-0.45em}%
                      \else{$''\hskip-0.45em$}\fi}
\def\tsp{\thinspace}

\title[The 421-day Periodicity in A0538--66]
      {On the Stability of the 421-day Periodicity in A0538--66}
\author[K.E. McGowan \& P.A. Charles]
{K.E. McGowan$^{1,2}$\thanks{email: mcgowan@lanl.gov}, P.A. Charles$^{2,3}$\\
$^{1}$Los Alamos National Laboratory, Los Alamos, NM 87545, USA\\
$^{2}$Department of Physics, University of Oxford, Oxford OX1 3RH\\
$^{3}$Department of Physics \& Astronomy, University of Southampton, 
Southampton, SO17 1BJ}

\date{Accepted 
      Received       }

\pagerange{\pageref{firstpage}--\pageref{lastpage}}
\pubyear{2002}

\begin{document}

\maketitle

\label{firstpage}

\begin{abstract}
{In this paper we analyse 70 years of archival Harvard and Schmidt
plate data of the 16.6~d Be X-ray binary A0538--66 in order to search for
the  presence of the long-term period of 420.82 $\pm$ 0.79~d found in
MACHO photometry (Alcock et al.\ 2001).  We find evidence for a
long-term period of 421.29 $\pm$ 0.95~d in the archival data, and
examine its stability.  We also combine the archival and  MACHO
datasets in order to improve the accuracy of the orbital period
determination, using a cycle-counting analysis to refine its value to
16.6460 $\pm$ 0.0004~d.  We also test the model proposed in our
previous paper (Alcock et al.\ 2001) with observations documented in
the literature for A0538--66 from 1980--1995, constraining the system
inclination to be $i>$74.9 $\pm$ 6.5\deg.}
\end{abstract}


\section{Introduction}
\label{sect:intro}

\ao\ is a recurrent X-ray transient in the LMC (White \& Carpenter
1978), which contains a neutron star (X-ray pulsations were detected
by the Einstein Observatory; Skinner et al.\ 1982) orbiting a bright
(B$\sim$15) Be star.  At various times the source exhibits 
He\tsp{\scriptsize II} $\lambda$4686 emission and P Cygni profiles
have been detected on Balmer and He\tsp{\scriptsize I} lines (e.g.\
Charles et al.\ 1983).  These spectral features  with such high
velocity components are not seen in classical Be stars.  The system
was found to have an orbital period of 16.6515 $\pm$ 0.0005~d from the
recurrence of its outbursts (Skinner 1981; hereafter S81), which are
interpreted as arising in a highly eccentric orbit (Charles et al.\ 1983).

In a previous paper (Alcock et al.\ 2001; hereafter Paper I) we
analysed \til5 years of optical monitoring of \ao\ by the MACHO
project.  This revealed a remarkably stable long-term modulation of
420.82 $\pm$ 0.79~d, together with the previously known 16.6510 $\pm$
0.0022~d period, hitherto interpreted as being orbital in origin.  A
model was presented in Paper I in which the origin of the long-term
period is due to the formation and depletion of an equatorial disc
surrounding the Be star.  In this model, as the disc forms, the
material shrouds the Be star and the brightness of the source is
reduced, as is observed in the light curve (see
Fig.~\ref{fig:ephem}, top panel; which is based on Fig.\ 1 of
Paper I).  The neutron star is then able to accrete from the
circumstellar matter and outbursts can occur.  The successive neutron
star passages deplete the available material and the source is seen to
brighten.  Once all the material is accreted, or ejected, the source
returns to a quiescent state in which the naked Be star can be
observed.

To investigate the periodicities found in Paper I we have
analysed archival photometric data of \ao\ from S81.  To
test our model we have compared the times of predicted outbursts with
observations of \ao\ from the literature.

\begin{figure*}
\begin{center}
\resizebox*{0.77\textwidth}{.25\textheight}{\rotatebox{-90}{\includegraphics{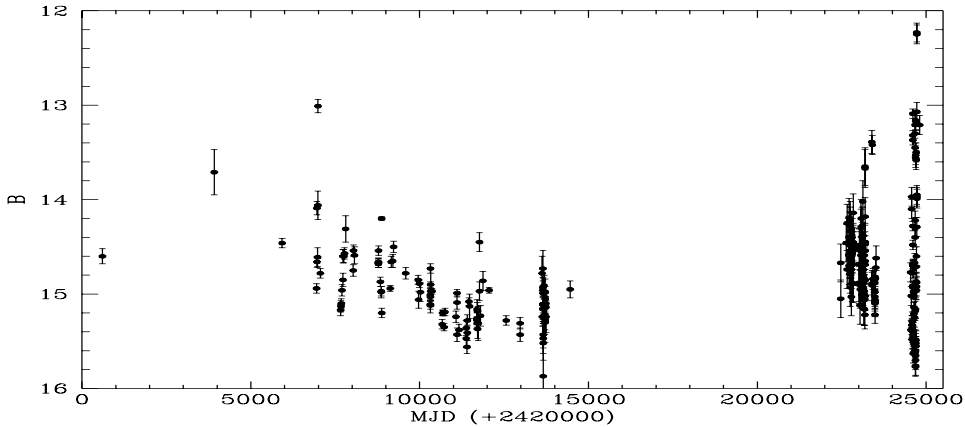}}}
\caption{Archival plate data (S81) showing the light curve of \ao\ from April 
1915 to July 1981.}\label{fig:skinner_lc}
\end{center} 
\end{figure*} 

\section{Archival Photographic Data}
\label{sect:harvard}

In 1981 Skinner analysed archival UK Schmidt and Harvard photographic
{\it B}-band plates of \ao, taken between 1915 and 1981.  From these
observations he derived the 16.6515~d orbital period, and an ephemeris
of T$_{\circ}$ = MJD~2443423.96, which has been the basis for studies
of \ao\ over the last 20 years.  Our MACHO observations revealed the
presence of the same orbital modulation with a period of 16.6510~d
(see Paper I), which within errors confirms Skinner's orbital period
for \ao.  It was noted that the outbursts observed in the MACHO light
curve are much smaller than those found by S81 and Densham et al.\
(1983).  Our aim in analysing the archival plate data was to
investigate both the presence and stability of this long-term period
over the much longer interval (\til70 years) of the archival
plate material.  Skinner's data is plotted in
Fig.~\ref{fig:skinner_lc}.

\subsection{The 421~d cycle}
\label{sect:har_421}

\begin{figure*}
\begin{center}
\resizebox*{.77\textwidth}{.25\textheight}{\rotatebox{-90}{\includegraphics{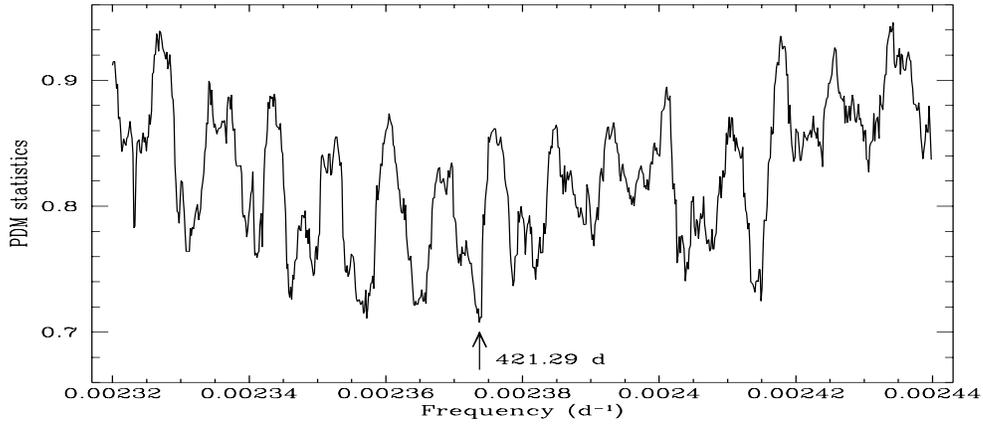}}}
\caption{Phase dispersion minimisation periodogram for the archival
data.  A frequency range of 2.32x$10^{-3}$--2.44x$10^{-3}$ cycle
d$^{-1}$ was searched with a resolution of 1.5x$10^{-7}$ cycle d$^{-1}$.}\label{fig:skinner_long_search}
\end{center} 
\end{figure*}

\begin{figure*}
\begin{center}
\resizebox*{.7\textwidth}{.55\textheight}{\rotatebox{0}{\includegraphics{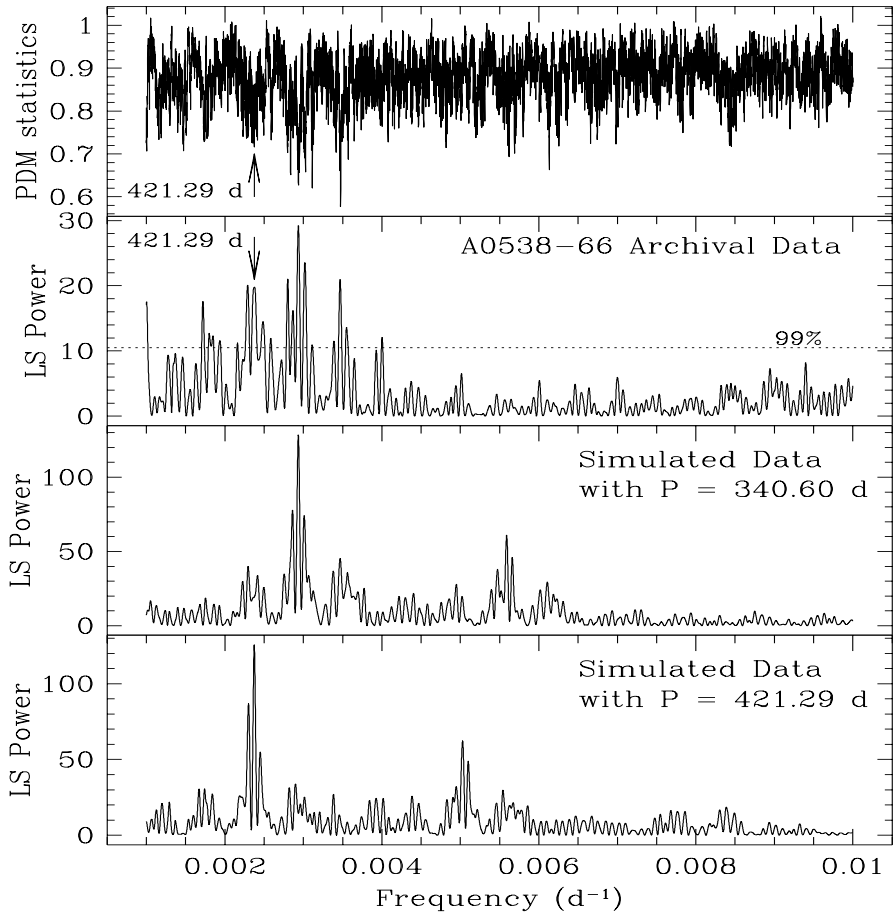}}}
\caption{Phase dispersion minimisation (first panel) and Lomb-Scargle
periodograms (second panel) for the archival data over a frequency
range of 0.001--0.01 cycle d$^{-1}$, with a resolution of 1x$10^{-6}$
cycle d$^{-1}$.  The third and fourth panels show the Lomb-Scargle
periodograms for the simulated light  curves, which contained a
sinusoidal signal of 340.60~d and 421.29~d,
respectively.}\label{fig:skinner_longer_search}
\end{center} 
\end{figure*}

To search for the long period found from the MACHO photometry we first
removed the 16.65~d periodic outburst events from the archival plate
data.  Using our {\it a priori} knowledge, we searched for a
periodicity near to 420.82~d.  We performed the temporal analysis
using a phase dispersion minimisation (PDM) periodogram (Stellingwerf
1978) over the frequency range 2.32x$10^{-3}$--2.44x$10^{-3}$ cycle
d$^{-1}$, with a resolution of 1.5x$10^{-7}$ cycle d$^{-1}$
(Fig.~\ref{fig:skinner_long_search}).  The lowest peak in the PDM
corresponds to P = 421.29~d, which is within the error for the MACHO
period.  We propagated an error for the peak in the PDM periodogram by
a Monte Carlo simulation in which we created artificial light curves
with the same mean and standard deviation as the archival light curve.
We calculated PDMs for each artificial dataset.  We then employed a
centroiding technique in which the mode of the peaks near the peak of
interest that are below a user defined threshold is calculated.  We
find P = 421.29 $\pm$ 0.95~d, which again encompasses the MACHO
period.  To assess the significance of this period, we performed a
period search over a longer frequency range, namely 0.001--0.01 cycle
d$^{-1}$, with a resolution of 1x$10^{-6}$ cycle d$^{-1}$
(Fig.~\ref{fig:skinner_longer_search}).  We employed a PDM
periodogram and Lomb-Scargle (LS) periodogram (Lomb 1976; Scargle
1982) for the temporal analysis.

The results from the longer search indicates that the 421.29~d period is
significant.  However, the LS periodogram has many peaks above the
99\% confidence level.  We created two simulated light curves using a
Gaussian random number generator with the same mean and standard
deviation as the archival data.  One contained an added sinusoidal
signal with a frequency equal to that of the highest peak in the LS
periodogram, while the other contained a signal corresponding to
421.29~d.  We then constructed a LS periodogram for each simulated
light curve, with the same frequency range and resolution as for the
longer search.  The peaks present in the LS periodogram for the
simulated light curves (Fig.~\ref{fig:skinner_longer_search},
third and fourth panels) show that the LS periodogram structure in the
real data cannot distinguish between the presence of one or two
periods.  It is also possible that the periodogram can be due to one
frequency which shifts slightly over the time span of the dataset.
The latter case is more likely given the nature of the long-term
periods in Be systems.  Hence we find marginal evidence, given the
poorer quality of the plate data, for a comparable periodicity to that
found in the MACHO data.

\begin{figure*}
\begin{center}
\resizebox*{.8\textwidth}{.25\textheight}{\rotatebox{-90}{\includegraphics{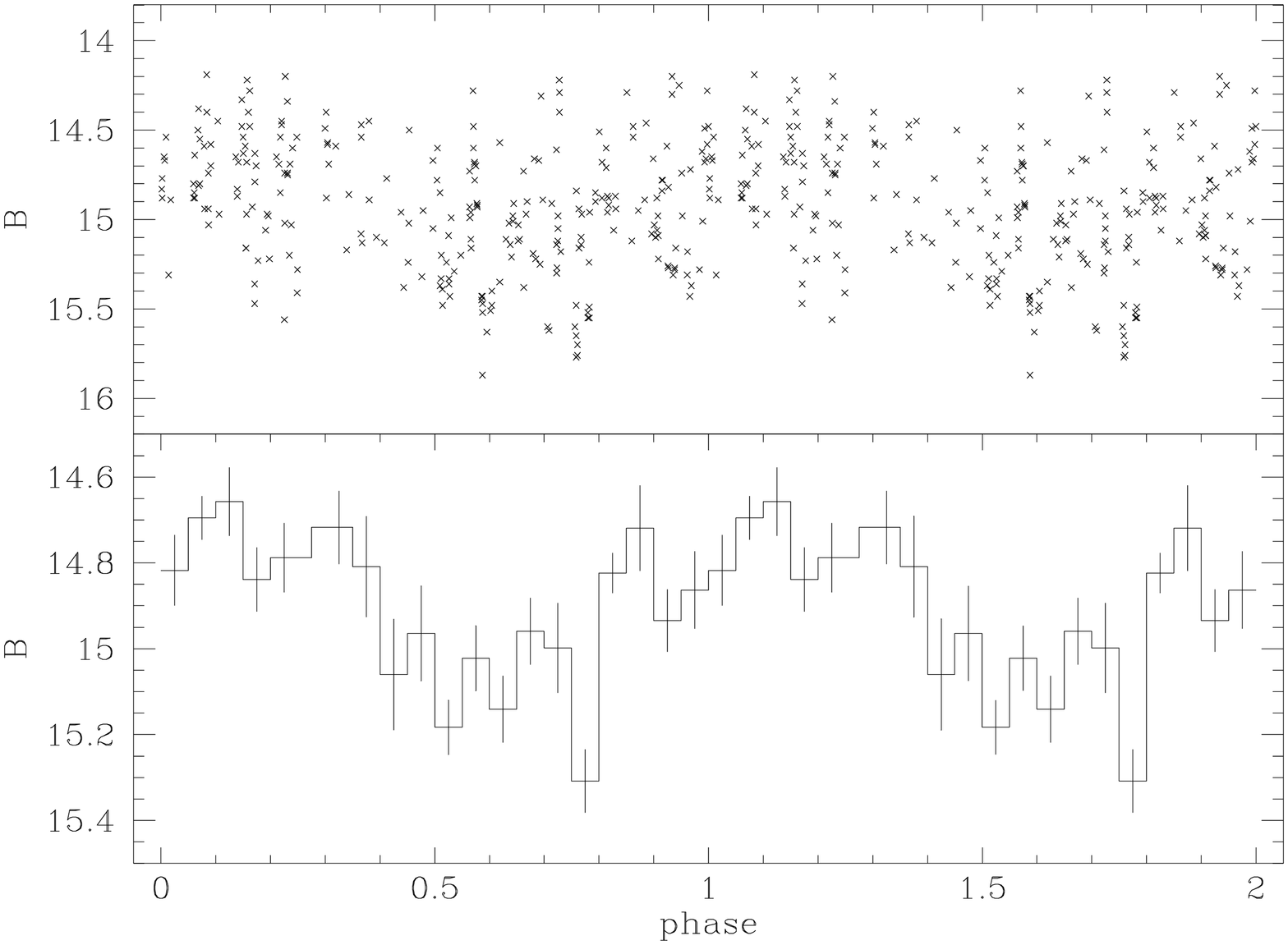}}}
\caption{Archival data folded on P = 421.29 d with T$_{\circ}$ = JD
2449002.109 (top), and in 20 phase bins (bottom).  Error bars for the
binned  light curve are the standard errors on the mean for the data
points in each bin.}\label{fig:skinner_long_fold}
\end{center} 
\end{figure*}

We folded the archival data on P = 421.29~d, using the MACHO ephemeris
T$_{\circ}$ = JD~2449002.109 (see Paper I), and then binned the data,
to investigate the form of the modulation
(Fig.~\ref{fig:skinner_long_fold}).  The folded and phase binned
light curves indicate that there is a periodic modulation present.
The form of the variability is marginally different to that for the
MACHO data, as the flat section of the folded light curve we obtained
for the MACHO data is not evident (see Fig.~\ref{fig:ephem}, top 
panel).  However, this may be due to there being a much greater
uncertainty in the {\it B} magnitudes determined from the archival
plates which leads to a larger scatter in the archival light curve.
We also note that a far greater number of cycles are used to create
the folded light curve of the archival data.  We find that the
modulation in the MACHO light curve varies from cycle to cycle, this
may also be occurring in the archival data, however the scatter would
mask this.

\subsection{The 16.6~d orbital period}
\label{sect:har_16}

We also investigated whether we could determine an improved orbital
period and ephemeris for our combined dataset spanning 83 years,
comprising the archival plate data and the MACHO data.  S81 found an
orbital period of 16.6515 $\pm$ 0.0005~d for the archival data and we
found P = 16.6510 $\pm$ 0.0022~d for the MACHO data.  Before we could
search for a common period we had to detrend the archival data; as for
the MACHO data, this was achieved by subtracting a third order
polynomial from the dataset.  We then combined these data with the
detrended MACHO data (see Paper I).

\begin{figure*}
\begin{center}
\resizebox*{0.77\textwidth}{.25\textheight}{\rotatebox{-90}{\includegraphics{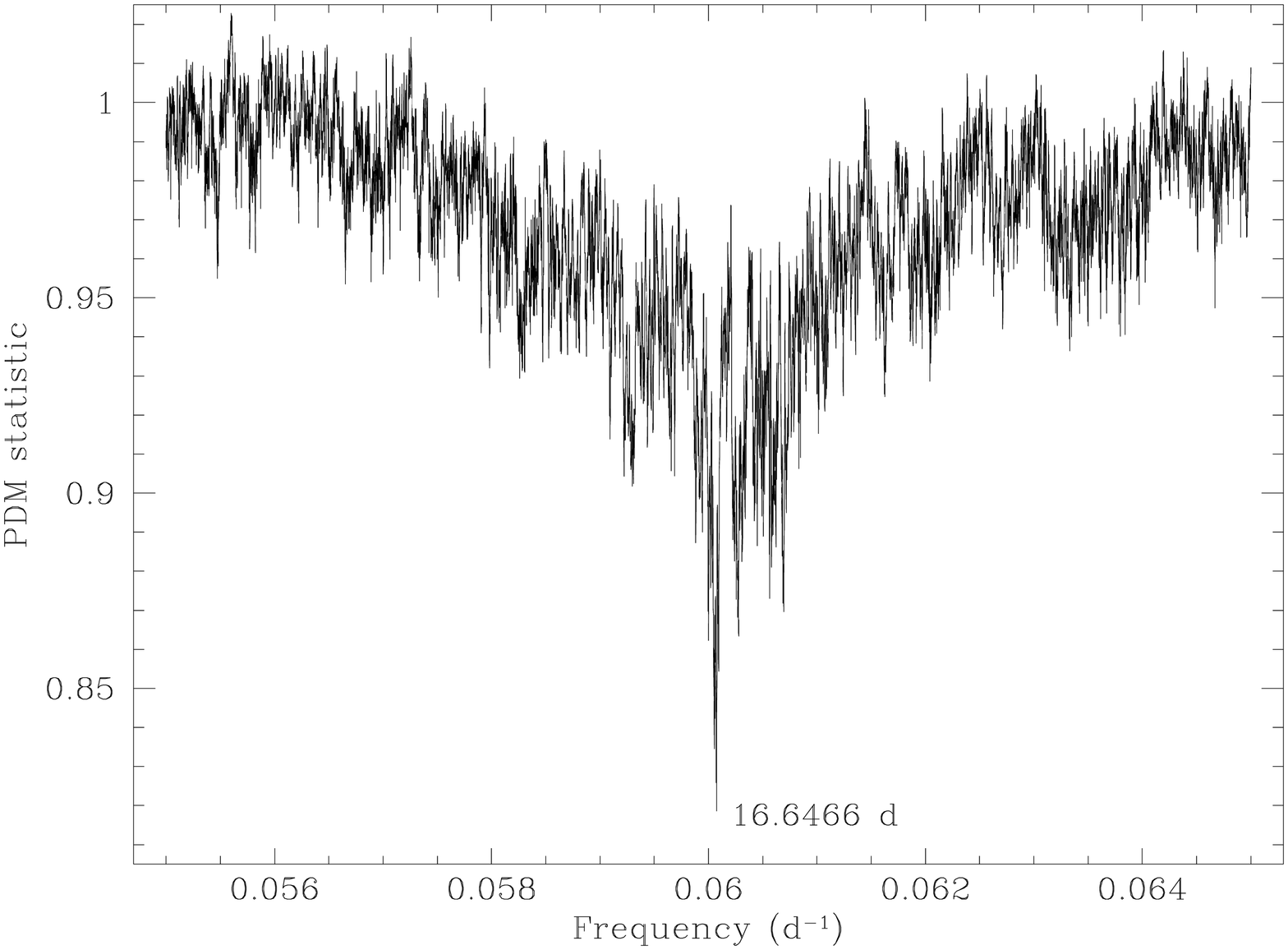}}}
\caption{Phase dispersion minimisation for the combined archival and
MACHO data. Frequency range 0.055--0.065 cycle d$^{-1}$ with resolution
8.25x$10^{-7}$ cycle d$^{-1}$.}\label{fig:skinner_short_search}
\end{center} 
\end{figure*} 

\begin{figure*}
\begin{center}
\resizebox*{0.7\textwidth}{.5\textheight}{\rotatebox{0}{\includegraphics{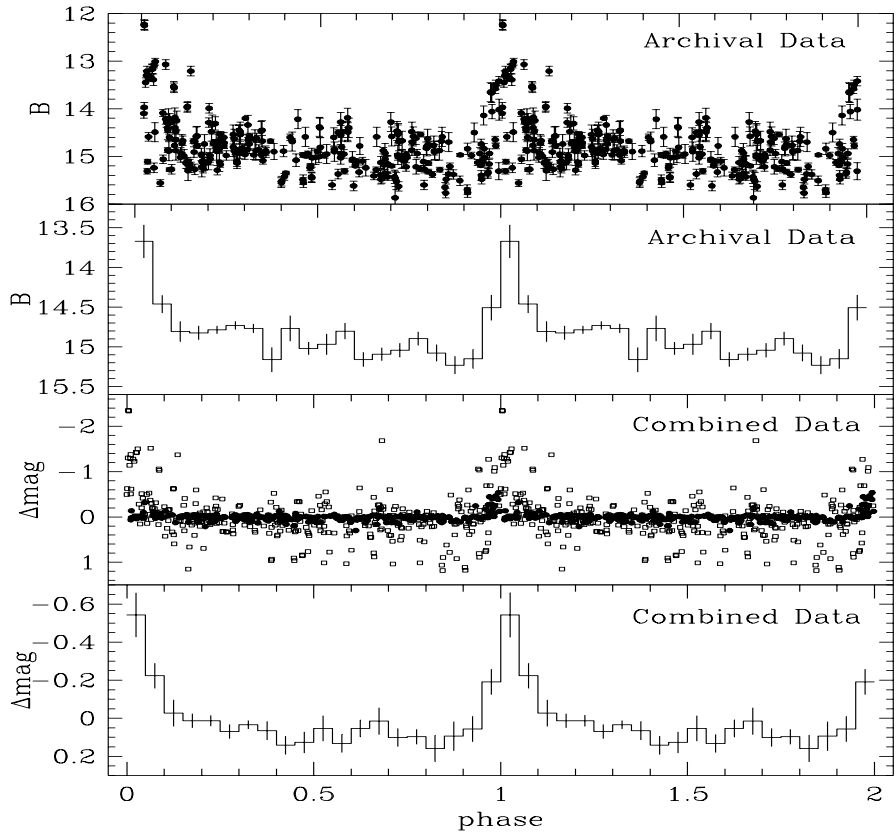}}}
\caption{Archival plate data (S81) folded on P = 16.6515~d using
T$_{\circ}$ = MJD~2443423.96 (first panel), and in 20 phase bins
(second panel).  Combined MACHO and archival \ao\ data folded on the
best-fit period of 16.6460~d with T$_{\circ}$ = MJD~2441443.143,
archival data are plotted as open squares, MACHO data are plotted as
filled circles (third panel), and in 20 phase bins (fourth panel).
Error bars for the binned light curve are the standard errors for the
data points in each bin.}\label{fig:pchi_fold}
\end{center} 
\end{figure*}

\begin{figure*}
\begin{center}
\resizebox*{0.8\textwidth}{.5\textheight}{\rotatebox{0}{\includegraphics{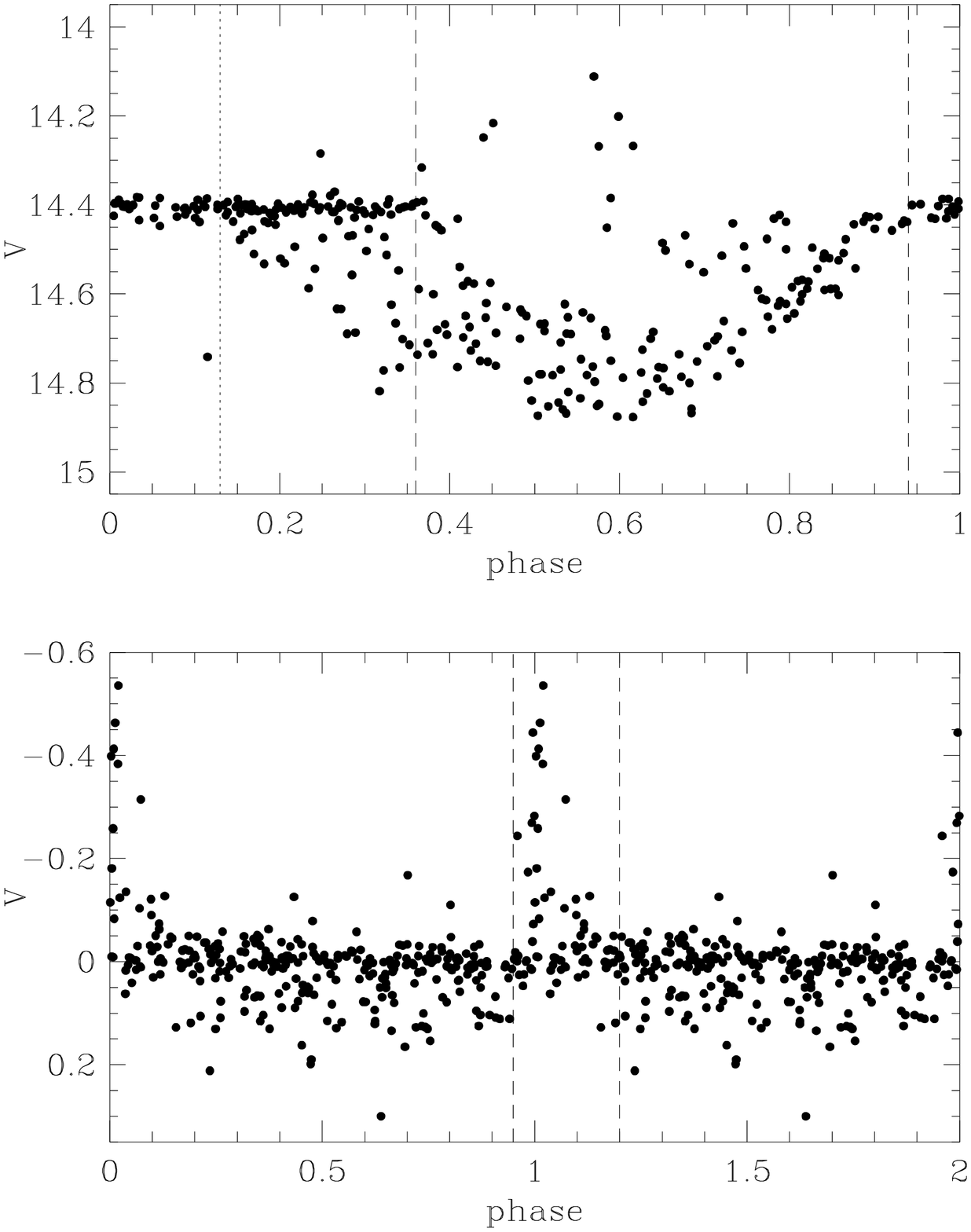}}}
\caption{The MACHO {\it V}-band light curve folded on P = 420.82 d
with  T$_{\circ}$ = JD~2449002.109 (top panel), and the detrended
light curve folded  on P = 16.6460~d with T$_{\circ}$ = MJD~2441443.1432
(bottom panel).  The dashed  lines indicate the phases of activity for
the 421 d cycle (top panel) and the  high state for the 16.6~d period
(bottom panel), the ranges are  $\phi_{421}$ = 0.36--0.94 and
$\phi_{16}$ = 0.96--0.2, respectively.  The  dotted line in the top
panel indicates the uncertainty for the lower  limit for the activity
phase in the 421 d cycle, and corresponds to  $\phi$ =
0.13.}\label{fig:ephem}
\end{center} 
\end{figure*}

Due to the highly non-sinusoidal nature of the outburst modulation
seen in both the archival data (see Fig.~\ref{fig:pchi_fold})
and the MACHO data (Fig.~\ref{fig:ephem}), a Fourier-based period
search is not appropriate.  A PDM was performed for a frequency range
0.055--0.065 cycle d$^{-1}$ with a resolution of 8.25x$10^{-7}$ cycle
d$^{-1}$, the resulting periodogram shows substantial fine structure
which makes determining the true period difficult
(Fig.~\ref{fig:skinner_short_search}).  The minimum value in the
PDM found using the centroiding technique, with an error propagated
with a Monte Carlo simulation as before, is P = 16.6466 $\pm$
0.0002~d.  The value for the period found from the combined archival
plate data and the MACHO data is lower than the periods found for each 
dataset alone (see S81 and Paper I, respectively).  The error found
also does not encompass the previous values.  We note that the range
of periods investigated in the combined search is narrower and the
resolution finer than the previous searches.

\begin{figure*}
\begin{center}
\resizebox*{.8\textwidth}{.32\textheight}{\rotatebox{-90}{\includegraphics{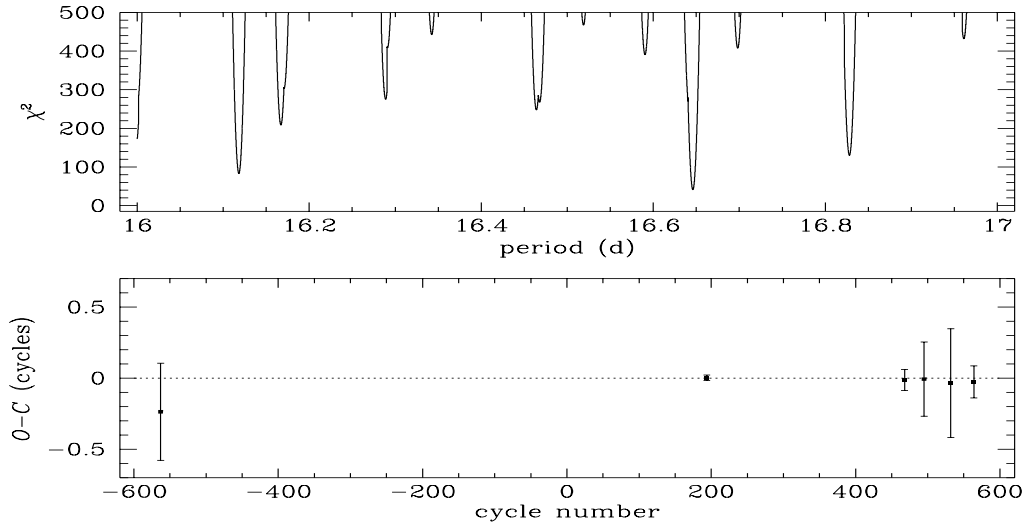}}}
\caption{$\chi^{2}$ for the predicted versus observed times of
outburst maximum, for a corresponding range of periods (top panel).
Within the range of possible periods consistent with the PDM result,
there is one minimum from the  cycle counting approach which has a
smaller $\chi^{2}$ than any others.   Bottom panel, for the best fit
period of 16.6460~d, the observed ({\it O}) --  computed ({\it C})
times of outburst maximum (for sections in which the  outburst peak
was present in both the folded and bin-folded data at a  significant
level only) are shown, in terms of cycles.}\label{fig:pchi}
\end{center} 
\end{figure*} 

To refine this result we attempted a cycle-counting analysis of the
outburst  timings.  We divided the combined data into 6 sections.  We
then took the  period given by S81 and folded and phase-binned each
section of data on it, using an arbitrary phase zero (T$_{\circ}$).
The phase of the peak of  the outburst ($\phi_{outburst}$) was
estimated with an error determined from  the other points in the peak.
The time for the outburst peak is given by
$T_{outburst,n}=T^{'}_{\circ,n}+\phi_{outburst}\times P_{outburst}$,
where
$T^{'}_{\circ,n}=T_{\circ}+int[(T_{midpt}-T_{\circ})/P_{outburst}]\times
P_{outburst}$, in order to give a time close to the mid-point time
($T_{midpt}$) of the {\it n}th dataset.  We are left with a set of 6
ephemerides spanning over 83 years, which should all agree with a
single  recurrence period.  For each cycle and trial period, a time of
minimum is  predicted ({\it C}) which is then compared to the observed
time ({\it O}), and  a simple $\chi^{2}$ minimisation for the {\it
O--C} values versus the trial  period then gives a best-fit period of
16.6460 $\pm$ 0.0004~d, and ephemeris T$_{\circ}$ = MJD~2441443.1432
(Fig.~\ref{fig:pchi}).  The {\it O--C} plot  for the best-fit
period is shown in Fig.~\ref{fig:pchi}.

The combined data were folded and phase-binned on the best-fit period
and  ephemeris derived from the cycle counting method
(Fig.~\ref{fig:pchi_fold}).   The value for the best-fit period is
again lower than the period found in S81 and Paper I.

\section{Archival Data from 1980--1995}
\label{sect:archive}

By studying previous observations of \ao\ from the literature we are
able to investigate our proposed model for the 421~d cycle.  \ao\ is
known to exhibit recurrent optical and X-ray outbursts on its orbital period.
By convention the source is described as being ``active'' when such
outbursts occur (which coincide with when the source is optically
bright).  The ``quiescent'' phase is taken to be when no
outbursts are detected, coinciding with when the source is optically faint.
Many observations of \ao\ have been taken near phase zero of the
16.6~d period hoping to detect outbursts, some have been successful, while
others not. 

In our model \ao\ is only in an active state, and hence outbursts can
occur, during the extended dip phase in the 421~d cycle, when the
system is faint.  The brightness of the source decreases due to the
formation of the Be star's equatorial disc, as the neutron star passes
through this disc it accretes material leading to outbursts.
Successive orbits of the neutron star depletes the circumstellar
matter causing the brightness of the source to increase.  \ao\ then
returns to quiescence when accretion, or ejection, of all the disc
material has occurred.  While the source is in quiescence it is seen
to be optically bright as we observe only the naked B star.  With our
model we can predict times when the source should display outbursts,
and when the bare B star can be observed.

It is clear then that the previously used {\it active/quiescent}
terminology is too crude to cover the range of behaviour that is
observed.  Therefore, using our model we have defined the following
activity states for the source :

\begin{itemize}
\item {\it ACTIVE} : in the 421~d cycle the source is only active during the
extended dip, i.e. $\phi_{421}$ = 0.36--0.94
\item {\it high} : in the active state, times of optical/X-ray outburst, or
mini-outburst, predicted by the 16.6~d orbital period, i.e. 
$\phi_{16}$ = 0.96--0.2 
\item {\it low} : times when outbursts should not be seen in the active state
\item {\it QUIESCENT} : times when only the naked B star should be
seen in the 421~d cycle, i.e. $\phi_{421}$ = 0.94--0.36
\end{itemize}
 
Fig.~\ref{fig:ephem} (top panel) shows the MACHO {\it V}-band light
curve folded on the 421~d period, the dashed lines indicate the lower
and upper limits for the active phase.  There is uncertainty in the
lower limit for the active phase in the 421~d cycle due to the
ill-defined ingress into the dip, which can be as low as $\phi_{421}$ = 0.13.  
The dotted line demonstrates the lowest limit that could be considered
as an active phase for the 421~d cycle.  The bottom panel of 
Fig.~\ref{fig:ephem} shows the detrended {\it V}-band light curve of
\ao\ folded on the best-fit 16.6~d period, the dashed lines indicate
the high state phase interval.

To see if our model agrees with observations we have taken published
data of \ao\ from 1980--1995 and calculated the phases of each
observation on the 421~d cycle.  We have also calculated the phases
for the observations using the newly determined best-fit period and
ephemeris, P = 16.6460~d and T$_{\circ}$ = MJD~2441443.1432, respectively.
A summary of the active and quiescent observations and how they relate
to our model is given below.  Details of the archival observations and
their phases, calculated with reference to our model, are shown in
Table 1 which is only available electronically via
http://www.raptor.lanl.gov/NoFrame\_Publications.htm.

\subsection{Active}
\subsubsection{High State}

Optical spectra of \ao\ taken during the active high state show a
number of features.  In general, the spectra are dominated by Balmer
and He\tsp{\scriptsize I} emission with P Cygni profiles (Charles
et al.\ 1983; Corbet et al.\ 1985; Corbet et al.\ 1997; Misselt,
Clayton \& Shulte-Ladbeck 1998).  The emission lines indicate the
presence of a disc.  At times of outburst maximum He\tsp{\scriptsize
II} $\lambda$4686, which is characteristic of X-ray binaries, is
observed (Charles et al.\ 1983; Hutchings et al.\ 1985; Corbet et al.\
1985; Menzies, Feast \& Howarth 1983).  The presence of
He\tsp{\scriptsize II} $\lambda$4686 emission just prior to outburst
maximum (i.e.\ $\phi_{16}$\geq 0.96) and post outburst maximum 
($\phi_{16}$\geq 0.1) is varied.  Corbet et al.\ (1985) do not detect 
He\tsp{\scriptsize II} emission in pre-outburst maximum spectra taken
in 1983 which agrees with observations of Charles et al.\ (1983),
however Corbet et al.\ (1985) found He\tsp{\scriptsize II} emission in
observations from 1982.  The spectra from 1983 also exhibit H and
He\tsp{\scriptsize I} absorption, while H$\beta$ is in emission
(Corbet et al.\ 1985).  He\tsp{\scriptsize II} is not found in spectra
taken after outburst maximum by Corbet et al.\ (1997) and Misselt et
al.\ (1998), but Charles et al.\ (1983) do detect it.

Howarth et al.\ (1984) described their UV spectra as ``outburst'',
``inter-outburst'' and ``quiescent''.  The distinction between
``outburst'' and ``inter-outburst'' was made based on the shape and
strength of the continuum, and the strength of the C\tsp{\scriptsize
IV} $\lambda$1550 emission.  The UV spectra taken during the active
high state classed as ``outburst'' spectra show strong, broad 
C\tsp{\scriptsize IV} $\lambda$1550, N\tsp{\scriptsize V}
$\lambda$1240 and He\tsp{\scriptsize II} $\lambda$1640 emission.  
The ``outburst'' spectra all have $0.0 < \phi_{16} < 0.1$.  The two
high state spectra classed as ``inter-outburst'' taken before outburst
maximum show C\tsp{\scriptsize IV} $\lambda$1550 and
N\tsp{\scriptsize V} $\lambda$1240, but no He\tsp{\scriptsize II}
$\lambda$1640 emission.  The emission features in the UV spectra
obtained by Charles et al.\ (1983) that coincided with an optical
outburst (B = 12.7--12.81) agree with the ``outburst'' spectra taken
by Howarth et al.\ (1984). 

{\it ROSAT} observations by Campana (1997) were described as quiescent
due to the low flux observed.  However, we find that the highest X-ray
luminosity measured of 1$\times 10^{36}$\ergsec\ occurs during an active
high state.  We suggest that this simply corresponds to a
low-amplitude outburst.  This is similar to the Corbet et al.\ (1997)
observations which they attribute to an outburst of low-amplitude.
Their active high state {\it ASCA} observation measured an average
X-ray luminosity of \til5.5$\times 10^{36}$\ergsec.

\subsubsection{Low State}

Active low state optical spectra display an early-type absorption
spectrum.  Emission lines of H, He\tsp{\scriptsize I} and
He\tsp{\scriptsize II} $\lambda$4686 are also detected (Murdin,
Branduardi-Raymont \& Parmar 1981; Charles et al.\ 1983; Corbet et al.\ 1985).
In addition to Balmer and He\tsp{\scriptsize I} absorption, Smale et al.\
(1984) find weak H$\beta$ and H$\gamma$ emission.  The lines display inverted
P Cygni profiles which vary, they suggest this is due to the infall of
material.  The presence of a disc is again inferred from the Balmer emission. 

The UV active low state spectra of \ao\ show no He\tsp{\scriptsize II}
$\lambda$1640 emission, but C\tsp{\scriptsize IV} $\lambda$1550 and
N\tsp{\scriptsize V} $\lambda$1240 emission are present (Howarth et
al.\ 1984).  The lines can aslo display P Cygni profiles (Charles et
al.\ 1983). 

{\it ROSAT} observations by Campana (1997) show that the X-ray
luminosity of \ao\ varies from 3$\times 10^{34}$\ergsec\ to
1$\times 10^{35}$\ergsec\ during the active low state.

\subsection{Quiescent}

Quiescent optical spectra are characterised by Balmer and 
He\tsp{\scriptsize I} absorption (Smale et al.\ 1984; Hutchings et
al.\ 1985; Misselt et al. 1998).  Hutchings et al.\ (1985) do not
detect any Balmer emission characteristic of a disc, however Smale et
al.\ (1984) find evidence for weak H$\beta$ and H$\gamma$ emission.
One observation of Misselt et al.\ (1998) shows only an absorption
spectrum, while the other shows an absorption spectrum with
H$\beta$ and He\tsp{\scriptsize II} $\lambda$4686 emission.  In our
model an absorption spectrum only should be observed during quiescence
as we are viewing the naked Be star.  The presence of emission lines
can possibly be explained by the uncertainty in the beginning of the
ingress into the extended dip, which can move the start of the active
phase to $\phi_{421}$ = 0.13.  Howarth et al.\ (1984) find very weak
or no C\tsp{\scriptsize IV} $\lambda$1550 emission in their quiescent
UV spectra.  Measurements of the X-ray luminosity of \ao\ in
quiescence yield values \leq1$\times 10^{35}$\ergsec.

\subsection{Photometry}

We find that a wide range of magnitudes are measured for \ao\ during
both active high and low states, and during quiescence.  The outbursts
recorded by Corbet et al.\ (1985) and van Paradijs et al.\ (1984), all
but one outburst observed by Charles et al.\ (1983) and one outburst
from Densham et al.\ (1984) fall in our active high state.  The rest
of the photometric observations occur during the active low state, or
quiescence.  The original classification of the active and quiescent
phases relied on the brightness of the source.  Our model finds that
the source is in its extended ``active'' phase when the source is
faint, hence we cannot be certain of the state due to the magnitude alone.
Generally, the lowest magnitudes observed for \ao\ are found to
coincide with our active low state which agrees with our model predictions. 

We note that Densham et al.\ (1983) recorded the brightest outbursts
of \ao\ at the end of 1981 and the beginning of 1982.  We find that
only one of these observations is during an active high state using our model.
In general however, our model seems to agree with the previously published
observations.  It seems that our model can fit outbursts of normal
intensity, but is unable to fit the unusually bright ones observed by
Densham et al.\ (1983).

\section{System Inclination}
\label{sect:model}

If we assume that the fading and brightening events in \ao's MACHO
light curve are due to the formation and depletion of the equatorial
disc around the Be star, we can make a simple estimate of a lower
limit for the inclination of the system.

Using the quoted range of spectral type (Charles et al.\ 1983) and
temperature (Allen 1973) for the Be star in \ao\ we can determine an
average effective temperature for the star.  Assuming a blackbody
spectrum at $\lambda$ = 550 nm (\til Johnston {\it V}), and
using the effective temperature above, we can calculate the
intensity that would be observed from a normal B star.  By taking the
observed difference between the 'flat' part and the dip minimum from
the MACHO {\it V}-band light curve (Fig.~\ref{fig:ephem}) we can
determine the intensity that would be observed at the minimum.  By
comparing the intensity calculated for a normal B star alone and the
intensity for the B star and disc, we can calculate the percentage
loss in intensity due to the disc at its maximum extent, assuming our
model is correct.

The Be star and disc can then be represented simply by a half circle
(for the unobscured hemisphere) and half an ellipse (for the obscured
hemisphere).  The semi-minor axis of the ellipse is related to the
inclination angle for the system, $b=R\cos i$, where $b$ is the
semi-minor axis of the ellipse, $R$ is the radius of the star, and $i$
is the inclination.

This assumes that the disc covers all of the star on the obscured
hemisphere at radii greater than the semi-minor axis, and that the
disc is not emitting any light.  This therefore sets a lower limit for
$i$, since if the disc were contributing to the observed magnitude,
the inclination would have to be higher to cover more of the star to
compensate for the light from the disc.  We find that the lower limit
for the inclination of \ao\ is 74.9 $\pm$ 6.5\deg.

\section{Conclusions} 
\label{sect:conc}

The orbital period for \ao\ was determined by S81 using archival
plates.  We re-analysed these data to investigate the stability of the
long-term period found in the MACHO data.  Given our {\it a priori}
knowledge, we find evidence for periodic variability on 
P = 421.29 $\pm$ 0.95~d for the archival data, which agrees with the
MACHO period (P = 420.82 $\pm$ 0.79~d), within errors.  This indicates
that the \til420~d cycle persists over a much longer interval than
presented in Paper I.  However, long-term monitoring of the source
must continue in order to answer this question convincingly.  

We also refined the orbital period and ephemeris by combining the
MACHO and plate data.  A straight-forward period search of the
combined data using a PDM periodogram led to a period of 16.6466
$\pm$ 0.0002~d.  This period is lower than those found by S81 
(P = 16.6515 $\pm$ 0.0005~d) and from the MACHO data 
(P = 16.6510 $\pm$ 0.0022~d).  We performed a cycle-counting analysis
to refine the period and found P = 16.6460 $\pm$ 0.0004~d, with an
ephemeris of T$_{\circ}$ = MJD~2441443.143.  The period found is
shifted by 0.0055~d from that of S81 using the archival plate data
alone, and by 0.0050~d from the period for the MACHO data.

We have studied published archival data from 1980--1995 to test our
model for  \ao.  Our results show that our model is in general
consistent with the  observations.  However, the published
observations largely correspond to our  inactive states for the
source.  A real test of our model is to observe an  outburst, or
mini-outburst, at a time predicted by our model.

By assuming that our model is correct i.e. the long-term
modulation is due to the formation and depletion of the Be star's
equatorial disc, and using the observed drop in intensity in the MACHO
{\it V}-band  light curve, we can determine a lower limit for the
inclination of the system  of $i$ = 74.9 $\pm$ 6.5\deg.

\section{Acknowledgements}

We thank Gerry Skinner for providing a copy of the archival Schmidt
and Harvard data.

\label{lastpage}

\end{document}